\documentclass[aps,prd,twocolumn,showpacs,amsmath,amssymb]{revtex4}
\usepackage{graphics}
\usepackage{graphicx}
\usepackage{latexsym}
\def\be{\begin{equation}}
\def\ee{\end{equation}}
\def\bi{\bibitem}
\begin{document}
\title{Inflation is the generic feature of phantom field-not the big-rip}
\author{Abhik Kumar Sanyal}
\email{sanyal_ak@yahoo.com}
\altaffiliation[Also at]{ RCRC, Dept. of Physics, Jadavpur University,
Kolkata-700032,India.}
\affiliation{Dept. of Physics, Jangipur College, Murshidabad,
West Bengal, India - 742213.} 
\begin{abstract}
A class of solutions for phantom field corresponding to a generalized k-essence
lagrangian has been presented, employing a simple method which provides the
scope to explore many such. All the solutions having dynamical state parameter
are found to touch the magic line $w=-1$ asymptotically. The solutions with
constant equation of state can represent phantom, quitessence or an ordinary
scalar field cosmologies depending on the choice of a couple of parameters of
the theory. For $w\approx-1$, quintessence and phantom models are
indistinguishable through the Hubble parameter. Finally, inflation rather than
big-rip has been found to be the generic feature of phantom cosmology.
\end{abstract}
\maketitle
\section{Introduction}
Current astronomical observations of luminosity distance-redshift relation, by  
standard candles such as supernova SN Ia, Galactic cluster measurements, 
particlarly by Sloan digital 
Sky survey, cosmic microwave background radiation with WMAP results and 
corresponding data analysis \cite{p} suggest that about 73\% of the total 
matter density of the Universe is in the form of a slowly varying dark 
energy with negative pressure, which is smoothly distributed all over the 
Universe with an equation of state $-1.45< w <-0.74$, at $95\%$ confidence 
level. In this connection Caldwell \cite{c} suggested a cosmological model 
with super-negative equation of state and dubbed it as phantom energy, which 
leads to the late time acceleration of the Universe. Such a field corresponds
to the action of an ordinary scalar field, with a reverse sign in the 
kinetic energy term. However, the solution presented in \cite{c}, found to
suffer from some serious set backs \cite{ck}. 
\par
\noindent
Other than instability, the main two problems associated with phantom type
fields in the cosmological context are, the big-rip, and the problem of
crossing the phantom divide line $w=-1$. Big-rip is a future (finite time)
singularity that replaces the big-bang singularity of Friedmann model, at which
all the cosmological parameters blow up, including curvature invariants.
However, soon Yurov \cite{av} has shown that the final state of phantom
cosmology may be inflation rather than big rip. Aref'eva et al \cite{af}
constructed a string inspired model whose solutions doesn't suffer
from big-rip and are stable under small fluctuations of the initial conditions
and special deviations of the form of the potential. Recently, Guo et al
\cite{z} have also demonstrated that the de-Sitter like solution is the late
time attractor of phantom cosmology. 
\par
\noindent
The second problem, on the other hand is much more serious. It has recently
been argued \cite{uv} that the best fit for the currently avilable data from
all the measurements mentioned at the beginning, is provided by a model of a
rapidly evolving dark energy from a dust-like $w=0$ at high redshift value
$z=1$ to a phantom-like $w<-1$, at the present. So, for a viable cosmological
model, the dark-energy equation of state must go over dynamically to $w=0$,
starting from below, $w < -1$, or vice versa. However, a dynamical transition
from $w<-1$ to $w>-1$, where, the equation of state corresponding to the
cosmological constant, $w=-1$, commonly known as phantom divide line, seems
to be impossible. This is because, matter with $w<-1$ violates dominant energy
condition and so such models are apparently supposed to suffer from
instability even in the classical level. Though the crossing of phantom divide
line is possible in some complicated multiple field theories, and non-minimally
coupled scalar field theories \cite{mf}, however, it has been proved \cite{pb}
that such crossing is impossible in a single field theory model, even for
generalized k-essence \cite{ams} non-canonical Lagrangian,
 \[ L = g(\phi)F(X) - V(\phi),\]
where, $X = \frac{1}{2} \partial_{\mu}\phi \partial^{\mu}\phi$, though 
Andrianov et al \cite{aaa} have claimed other way round. 
In any case, if we believe in an equation of state less than $-1$, we have to 
deal with phantom models seriously, since nothing other than phantom can live 
on the other side of the barrier. Following the above disussions, we are thus
motivated to explore the cosmological behaviour of the phantom field, through 
mostly a new class of an exact solutions of a generalized single field phantom 
model. In the process we shall show that, though it is indeed impossible to 
cross the phantom divide line by a single field model, however, all the 
solutions corresponding to dynamical equation of state parameter, 
asymptotically touch and others can be fit with, the magic line $w=-1$. This 
proves inflation, rather than big-rip is the generic feature of phantom 
cosmology. Further, it has been shown that the solutions with constant equation 
of state represent all the phantom, quintessence and ordinary scalar field 
cosmological models depending on the the choice of a couple of parameters of 
the theory. For $w\approx-1$, the quintessence and the phantom models are found 
to be indistinguishable through the asymptote of the Hubble parameter.
\par
\noindent
In the following section we write the action and the field equations of a 
generalized k-essence Lagrangian and develop a rather straight forward method 
to extract a class of solutions. In section 3, we present the set of solutions 
for different form of the potential.  
\section{Action and the field equations}
We start with a single field generalized k-essence action \cite{ams} 
(see also Vikman in \cite{pb}) which can be expressed in its simplest form as,
\be
S=\int d^4 x\sqrt{-g}[\frac{\kappa^2}{2} 
R+\frac{g(\phi)}{2}\phi,_{\mu}\phi^{,\mu}-V(\phi)],
\ee
where, we dub $g(\phi)$ as the k-essence parameter, which has got a Brans-Dicke 
origin, $g=\frac{\omega}{\phi}$ too, $\omega(\phi)$ being the Brans-Dicke 
parameter, though the coupling here is minimal. Note that the sign of the 
kinetic energy term depends on the k-essence parameter for a real scalar field 
$\phi$. For a positive potential $V(\phi)$, the action (1) represents that for 
phantom field if $g(\phi)$ is positive, while it turns to ordinary scalar field 
action for negative $g(\phi)$. Since phantom field has originated from string 
field theory, so the dimensionless coupling constant $\kappa^2$ is related to 
the string and the reduced Planck masses. We have chosen $\kappa^2 = 1$, for 
all practical purpose corresponding to the standard cosmological units 
$8\pi G = c =1$. For spatially flat, homogeneous and isotropic Robertson-Walker
space-time $k=0$, 
\[
ds^2 = -dt^2 + a^2(t)[dr^2+r^2 d\theta^2+r^2 sin\theta^2 d\phi^2],
\]
the field equations in terms of the Hubble parameter $H=\dot a/a$ are
\be
2\dot{H}+3H^2 = \frac{g}{2}\dot{\phi^2}+V(\phi)= - p,
\ee
\be
\ddot{\phi}+
3H\dot{\phi}+\frac{\dot{\phi^2}}{2}\frac{g'}{g}-\frac{V'}{g}=0,
\ee
\be
3H^2=-\frac{1}{2}g\dot{\phi^2}+V(\phi)=\rho,
\ee
where, dash $(')$ stands for derivative with respect to the scalar field 
$\phi$, while $p$ and $\rho$ are the pressure and the energy density of the
phantom field respectively. Now, differentiating equation (4) we get,
\[
g\phi\ddot{\phi}=-6H\dot H-\frac{1}{2}g'\dot{\phi}^3+V'\dot{\phi}.
\]
So, eliminating $\ddot{\phi}$ between the above equation and equations (3)
we obtain,
\be
\dot \phi= 2\frac{H'}{g}.
\ee
Note that in the above equation $H$ has been expressed as a function of
$\phi$ instead of time, which is referred to as the superpotential 
$H(t)=W(\phi(t))$ for a single field \cite{sp}. However, we do not use this 
notation, instead we use, $H(t) = H(\phi(t))$. It is possible to write yet 
another equation by eliminating $\dot \phi$ between equations (4) and (5), 
viz.,
\be
2 H'^2+3 g H^2 = g V(\phi).
\ee
So, (5) and (6) are the two equations that we need to solve for the Hubble
parameter $H$ and correspondingly the scale factor $a$ and also for the phantom 
field $\phi$. This is achieved only if $g(\phi)$ and the phantom field 
potential $V(\phi)$ are known. Not being sceptic, one can choose $\dot \phi$ as 
a function of $\phi$ so that equation (5) gets integrated immediately and the 
Hubble parameter is expressed in terms of the field $\phi$, rather than 
choosing the so called super-potential as a polynomial in $\phi$, restricting 
it to not more than third degree \cite{af}. In this method one can explore 
variety of forms for the so called super-potential, including inverse power law 
and the exponential ones. Then for different choice of potential $V(\phi)$, it 
is possible to extract a class of exact solutions of equation (6).
\par
\noindent
In the following section we have explored a class of exact solutions choosing  
in each of the three subsections different functional form of $\dot\phi$ and 
correspondingly different form of the potential $V(\phi)$.     
\section{Some exact solutions and the behaviour of the phantom field}
As already mentioned we shall have to solve equations (5) and (6) for $H, \phi,
g(\phi)$ and $V(\phi)$, which is not possible unless two are known. In the
following three subsections we have chosen three different forms of $g(\phi)$ 
in such a way that equation (5) gets integrated immediately. Then for different
choice of the potential $V(\phi)$ mostly a new class of exact solutions 
has been excavated and corresponding behaviour of the phantom field has been 
studied.
\subsection{\underline{Let, $g\dot{\phi} = n$ - a constant.}}
Under the above choice equation (5) can be integrated to yield,
\be
H=\frac{n}{2}\phi,
\ee
which is devoid of a constant of integration. Thus, this situation corresponds
to the linear functional dependence of the so called superpotential \cite{sp} 
on the field $\phi$. Hence, equation (6) now takes the form,
\be
\dot \phi+\frac{3n}{2}\phi^2-\frac{2V(\phi)}{n}=0,
\ee
as a result, the k-esence parameter $g(\phi)$, the kinetic energy $(K)$, the 
energy-density $(\rho)$, the pressure $(p)$ of the phantom field and hence 
the equation of state $w$, under consideration take the following form,
\[
g(\phi)=\frac{n}{\dot \phi}=\frac{2n^2}{(4V(\phi)-
3n^2\phi^2)};\]
\be
K=-\frac{1}{2}g\dot{\phi}^2 =-\frac{1}{4}(4V(\phi)-3n^2\phi^2);
\rho= \frac{3n^2}{4}\phi^2;
\ee
\[
p= \frac{3n^2}{4}\phi^2-2V(\phi);
\;\;
w= \frac{p}{\rho}= 1-\frac{8V(\phi)}{3n^2\phi^2}.
\]
Now, under some suitable choice of the potential $V(\phi)$ one can solve for
the scale factor and the field variable in view of the above two equations (7)
and (8), and can study the behaviour of all other parameters of the theory
from equation (9), which we carry out in the following.\\

\noindent
{\underline{{Case 1}, Let, $V = V_{0}$, a constant}}.\\

\noindent
For the above choice of a constant potential the action (1) now takes the 
following form,
\be
S=\int d^4 x\sqrt{-g}[R+\frac{n^2}{4V_{0}-3n^2\phi^2}\phi,_{\mu}\phi^{,\mu}
-V_{0}].
\ee
Equation (8) can now be integrated to yield,
\be
\phi=\frac{2}{n}\sqrt{\frac{V_{0}}{3}}tanh[\sqrt{3V_{0}}(t-t_{0})].
\ee
Hence, equation (7) can be solved for the scale factor as,
\be
a=a_{0}\sqrt[3]cosh[\sqrt{3V_{0}}(t-t_{0})].
\ee
The deceleration parameter is
\be
q=-\frac{\ddot{a}/a}{\dot{a}^2/a^2}= -3coth^2[\sqrt{3V_{0}}(t-t_{0})]+2.
\ee
The so called k-essence parameter, $g$, is found to have 
the following form,
\be
g=\frac{n^2}{2V_{0}}cosh^2[\sqrt{3V_{0}}(t-t_{0})].
\ee
Finally, the kinetic energy, the pressure, the energy-density and the equation 
of state of the phantom field under consideration are evaluated as,
\[
K=-V_{0}sech^2[\sqrt{3V_{0}}(t-t_{0})];\]
\be
p= - V_{0}\{2-tanh^2[\sqrt{3V_{0}}(t-t_{0})]\};
\ee
\[
\rho= V_{0}tanh^2[\sqrt{3V_{0}}(t-t_{0})];\;
w=1-2coth^2[\sqrt{3V_{0}}(t-t_{0})].
\]
Thus the above solutions depict that the cosmic evolution started from a
constant value of the scale factor which grew exponentially. The phantom field 
which was created as the Universe started evolving, asymptotically settles down 
to a constant value $\frac{2}{n}\sqrt{\frac{V_{0}}{3}}$, while the Hubble 
parameter is zero initially and is $\frac{V_{0}}{3}$ at the end. The 
deceleration parameter guarantees cosmic acceleration, which falls off and ends 
up with a value $-1$ asymptotically. The k-essence parameter, $g(\phi)$ starts 
from a constant value $\frac{n^2}{2V_{0}}$, grows indefinitely without a flip 
in sign, while the  kinetic energy of the phantom field starts from $-V_{0}$, 
increases with the evolution and asymptotically vanishes, demonstrating that 
such model ends up with a bare cosmological constant and is not viable of 
crossing the phantom divide line. The equation of state which starts from a 
large negative value and finally settles down to $-1$, also depicts the same 
feature. The energy density of the model increases from zero value and ends up 
at $V_{0}$. In this simple model, with it's natural cut off, the energy density 
never grows large enough to tear off gravitationally bound objects. Thus, this 
model never encounters big-rip rather it clearly indicates inflation with $p = 
-\rho,$ corresponding to a bare cosmological constant $V_{0}$ at the end
\cite{av}. \\

\noindent
{\underline{{Case 2}, Let, $V = V_{0}\phi^2$}}.\\

\noindent
Under the above asumption, $w$ becomes nondynamical, as can be seen in view of 
equation (9), so there does not arise any question of the crossing. The action in this case reads,
\be
S=\int d^4 
x\sqrt{-g}[\frac{R}{2}+\frac{n^2}{(4V_{0}-3n^2)\phi^2}\phi,_{\mu}\phi^{,\mu}
-V_{0}\phi^2].
\ee
Following the same above procedure, $\phi$ can be evaluated as,
\[
\phi= -\frac{2n}{4V_{0}-3n^2}(t-t_{0})^{-1}.
\]
The action (16) guarantees that the field is phantom only if $4V_{0}>3n^2$,
which indicates that big-rip can clearly be avoided for $n<0 =-m^2$. Hence, 
the complete set of solutions is found as, 
\[
\phi= \frac{2m^2}{4V_{0}-3m^4}(t-t_{0})^{-1};\;\;\;\;
a=a_{0}(t-t_{0})^{-\frac{m^4}{4V_{0}-3m^4}};
\]
\[
q=-1-\frac{4V_{0}-3m^4}{m^4};\;\;
g=\frac{4V_{0}-3m^4}{2}(t-t_{0})^2;
\]
\be
K=-\frac{m^4}{(4V_{0}-3m^4)(t-t_{0})^{2}};
\rho=\frac{3m^8}{(4V_{0}-3m^4)^2(t-t_{0})^{2}};
\ee
\[
p=-m^4\left(\frac{8V_{0}-3m^4}{(4V_{0}-3m^4)^2}\right)\frac{1}{(t-t_{0})^{2}};
w=-\frac{8V_{0}-3m^4}{3m^4}.
\]
\begin{figure}
\includegraphics{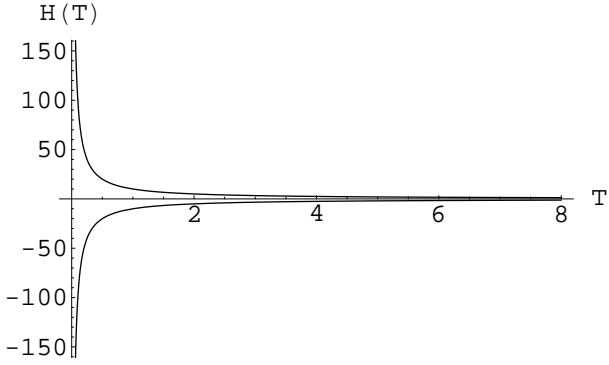}
\caption{Hubble parameters $H(T)$ admitted by solutions (29) corresponding to 
the phantom field ($n=-1$, lower curve) and the quintessence field ($n=1$, 
upper curve), is plotted against $(T)$, where, $T=(t-t_{0})$ and choosing 
$|4V_{0}-3n^2|=0.1$, so that $w=-1.06$ (phantom) and $w=-0.93$ (quintessence).
\label{figure1}}
\end{figure}
Thus the field vanishes asymptotically, while the k-essence parameter grows 
indefinitely starting from zero. The energy density also vanishes 
asymptotically, 
starting from an indefinitely large value, while the kinetic energy and the 
pressure acquire the same fate starting from a large negative value. Thus at 
the end there is no dark energy at all and we are left with vacuum Einstein's
equation. However, the problem encountered in this model as mentoned earlier, 
is that the deceleration parameter and the equation of state do not evolve with 
time. Nevertheless, $V_{0}$ can be chosen close enough to $\frac{3m^2}{4}$ in 
order to fit the present observable constraint on the equation of state 
parameter, which is pretty close to one (see, eg., Melchiorri et al in 
\cite{p}). It is interesting to note that the same set of solutions represents 
a model for quintessence field \cite{q} provided, $n>0$ and 
$4V_{0}<3n^2=3m^4<8V_{0}$, 
which can be a best fit with the magic line $w=-1$ from above, if 
$3n^2\approx4V_{0}$. Finally, the solutions represent a model for an ordinary 
scalar field too, with $w>0$ provided, $n>0$ and $3n^2=3m^4>8V_{0}$. No one yet 
knows whether the state parameter is dynamical. If it is not, then this is the 
"three-in-one" situation that covers the whole picture which are of course 
mutually exclusive. It can be mentioned at this juncture that a form invariance 
transformation \cite{cl} leads to phantom cosmology, starting from scalar field 
theories. We have been able to show that such transformation corresponds simply 
to adjustant of a couple of parameters of the theory. In figure 1. we have 
shown how the Hubble parameter evolves in time, both in the phantom and 
quintessence model. It demonstrates that for an equation of state close enough 
to $-1$, it is impossible to identify between the two. However, since the 
Hubble parameter vanishes asymptotically for both the models, so either we are 
living at the early stage of evolution or we have to discard this model.

\subsection{\underline{Let, $g\dot{\phi} = n\phi$.}}
Under the above choice, where, $n$ is a constant, equation (5) can be
integrated to yield,
\be
 H= \frac{n}{4}\phi^2,
\ee
ignoring the constant of integration. Now in view of equation (18), equation 
(6) takes the following form,
\be
\phi\dot \phi+\frac{3n}{8}\phi^4 -\frac{2}{n}V(\phi) = 0.
\ee
Accordingly, the k-essence parameter, the kinetic energy, the pressure, the 
energy-density of the phantom field and finally the equation of state take the 
following forms,
\[g=\frac{8n^2\phi^2}{16V(\phi)-3n^2\phi^4}
;K=-[V-\frac{3n^2}{16}\phi^4];
\]
\be
p= \frac{3n^2}{16}\phi^4 - 2V(\phi);
\rho=\frac{3n^2}{16}\phi^4;\;
w=1-\frac{32}{3n^2}\frac{V(\phi)}{\phi^4}.
\ee
Now, we proceed as before to generate solutions of the field equations (18) and
(19) under some suitable choice of the potential $V(\phi)$ , which we carry out
in the following.\\

\noindent
{\underline{{Case 1}, Let, $V = V_{0}$, a constant}}.\\

\noindent
For a constant potential the action (1) now takes the following form,
\be
S=\int d^4 x\sqrt{-g}[R+\frac{4n^2\phi^2}{16V_{0}-3n^2\phi^4}
\phi,_{\mu}\phi^{,\mu}-V_{0}].
\ee
As before the field variables can be evaluated in view of the equations (18)
and (19) as,
\[
\phi=2\sqrt[4](V_{0}/3n^2)(\sqrt{tanh[\sqrt{3V_{0}}(t-t_{0})]});\]
\be
a=a_{0}\;\sqrt[3]cosh[\sqrt{3V_{0}}(t-t_{0})].
\ee
All other parameters are now found in view of equation (20) as follows,
\[
q=-3coth^2[\sqrt{3V_{0}}(t-t_{0})]+2;
g =\frac{2ntanh[\sqrt{3V_{0}}
(t-t_{0})]}{\sqrt{3V_{0}}sech^2[\sqrt{3V_{0}}(t-t_{0})]};\]
\be
K=- V_{0} sech^2[\sqrt{3V_{0}}(t-t_{0})];
\rho=V_{o}tanh^2 [\sqrt{3V_{0}}(t-t_{0})];
\ee
\[
p=-V_{0}(2-tanh^2 [\sqrt{3V_{0}}(t-t_{0})]);
w=1-2coth^2[\sqrt{3V_{0}}(t-t_{0}].
\]
These solutions are almost identical to those obtained in case 1 of subsection 
A. The only difference is that the phantom field asymptotically settles at a 
slightly different value, $2\sqrt[4]{V_{0}/3n^2}$, While, the k-essence 
parameter starts from zero instead of a finite value, but grows indefinitely, 
as before. So, it appears that, as long as the form of the potential 
remains the same, the form of the Hubble parameter does not make any 
appreciable difference in the solutions. Even for $H\propto\phi^3$, or 
$H\propto\phi^{-1}$, it has been checked that (not shown here) the solutions do 
not differ. \\

\noindent
{\underline{{Case 2}, Let, $V = V_{0}\phi^2$}}.\\

\noindent
For such a quadratic form of the potential the action (1) can be expressed as,
\be
S=\int d^4 x\sqrt{-g}[\frac{R}{2}+\frac{4n^2}{(16V_{0}-3n^2\phi^2)}\phi,_{\mu}\phi^{,\mu}
-V_{0}\phi^2].
\ee
In view of equations (18) and (19) the field variables in this case are found
as,
\[
\phi=\sqrt{\frac{2V_{0}}{n}}\frac{e^{\frac{2V_{0}}{n}(t-t_{0})}}{\sqrt{1+
\frac{3n}{8}e^{4\frac{V_{0}}{n}(t-t_{0})}}};\]
\be
a=a_{0}[1+\frac{3n}{8}e^{\frac{4V_{0}}{n}(t-t_{0})}]^{\frac{1}{3}}.
\ee
All the parameters of the theory are obtained in view of equation (20) as,
\[
q=-[1+\frac{8}{n}e^{-\frac{4V_{0}}{n}(t-t_{0})}];
g=\frac{n^2}{2V_{0}}[{1+\frac{3n}{8}e^{\frac{4V_{0}}{n}(t-t_{0})}}];\]
\[
K=-\frac{2V_{0}^2 e^{\frac{4V_{0}}{n}(t-t_{0})}}
{n[1+\frac{3n}{8} e^{\frac{4V_{0}}{n}(t-t_{0})}]^{2}};
\rho=\frac{3V_{0}^2\; e^{\frac{8V_{0}}{n}(t-t_{0})}}{4
[1+\frac{3n}{8} e^{\frac{4V_{0}}{n}(t-t_{0})}]
^{2}};
\]
\be
p=-\frac{2V_{0}^2}{n} e^{\frac{4V_{0}}{n}(t-t_{0})}
\left[\frac{2+\frac{3n}{8}e^{\frac{4V_{0}}{n}(t-t_{0})}}{(1+\frac{3n}{8}
e^{\frac{4V_{0}}{n}(t-t_{0})})^2}\right];
\ee
\[
w=-[1+\frac{16}{3n}e^{-\frac{4V_{0}}{n}(t-t_{0})}].
\]
The Universe here again undergoes an exponential expansion. The field variable
$\phi$ grows, starting from a constant value rather than zero and 
asymptotically 
becomes yet another constant. The deceleration parameter starts with a value 
$<-1$ but finally settles down to $-1$, while the k-essence parameter also 
starts from a constant value and ultimately becomes infinitely large. The 
prssure as well as the kinetic energy begin with negative values and while the 
former ends up with less negative value $-\frac{16V_{0}^2}{3n^2}$, the later 
vanishes at the end. The energy density grows and finally reaches a moderate 
value $\frac{16V_{0}^2}{3n^2}$, and so, $\rho+p = 0$, at the end. The equation 
of state starts with a value much less than $-1$ and at the end settles down to 
$-1$, indicating inflation \cite{av}, rather than big-rip. It is noticeable 
that under no circumstances $n$ can be made negative, and so quintessence or 
the ordinary scalar field solutions are not realizable here, as was in earlier 
situation with quadratic potential. Thus the form of the Hubble parameter 
indeed has got a role towards the nature of the solutions.\\

\noindent
{\underline{{Case 3}, Let, $V = V_{0}\phi^4$}}.\\

\noindent
The above choice of quartic potential leads to the following form of the action (1),
\be
S=\int d^4 x\sqrt{-g}[\frac{R}{2}+\frac{4n^2}{\phi^2(16V_{0}-3n^2)}\phi,_{\mu}\phi^{,\mu}
-V_{0}\phi^4].
\ee
Further, equations (18) and (19) are used to determine the evolution of the
phantom field and the scale factor as,
\[
\phi=
\sqrt{\frac{-4n}{(16V_{0}-3n^2)(t-t_{0})}}\;=
\frac{2m}{\sqrt{(16V_{0}-3m^4)(t-t_{0})}};
\]
\be
a=a_{0}(t-t_{0})^{-(\frac{m^4}{16V_{0}-3m^4})}.
\ee
where, we have made $n$ negative $(n=-m^2)$ . Rest of the parameters are found 
in view of equation (20) as,
\[
q=-\frac{16V_{0}-2m^4}{m^4};
K=-\frac{m^4}{(16V_{0}-3m^4)(t-t_{0})^{2}};
\]
\be
g=2m^2(t-t_{0});\rho=\frac{3m^8}{(16V_{0}-3m^4)^{2}(t-t_{0})^{2}};
\ee
\[
p=-[\frac{32V_{0}-3m^4}{(16V_{0}-3m^4)^{2}}]\frac{m^4}{(t-t_{0})^{2}};
w=-\frac{(32V_{0}-3m^4)}{3m^4}.
\]
So, this solution behaves almost identically to that we obtained with quadratic 
potential (case 2 of subsection A). The Universe undergoes constant 
acceleration with constant equation of state for the phantom field which can be 
tuned close to $1$ to meet with the present observable constraint (see, eg., 
Melchiorri et al in \cite{p}). The phantom field and the energy-density both 
start from an indefinitely large value but vanish at the end, while the 
pressure and the kinetic energy encounter the same fate starting from a large 
negative value. It should be noted that one requires $(16V_{0}-3m^4)>0$ to 
ensure action (27) admits phantom field. As observed in case 2 of previous 
subsection, here again the solutions represent those for a quinessence field 
\cite{q}
for $n>0$ provided, $3n^2>16V_{0}$. Note that $3n^2\approx16V_{0}$ may be the 
best fit with the magic line from both ends. The above solutions represent  
an ordinary scalar field too, provided, $3n^2>32V_{0}$. Thus this situation is 
identical to that obtained earlier (case 2 of subsection A). Hence the same 
figure 1. represents the evolution of the Hubble parameter, considering, 
$|16V_{0}-3n^2|=0.1$, instead. So the same cosmological model under different 
choice of the parameters leads to all the three regimes, viz., positive 
definite, $w\ge0$, negative  $-1<w<0$ and supernegative, $w<-1$ equation of 
states.
\subsection{\underline{Let, $g\dot{\phi} = 2 n e^{l\phi}$.}}
Under such assumption equation (5) is integrated and we find,
\be
H=\frac{n e^{l\phi}}{l},
\ee
which indicates that the so called superpotential \cite{sp} has exponential 
functional dependence on $\phi$, in sharp contrast to Aref$^,$eva et al 
\cite{af}, who restricted it to a polynomial in $\phi$, allowing not more 
than third degree. Equation (6) can now be expressed as,
\be
n l^2e^{l\phi}\dot \phi+ 3 n^2 e^{2 l\phi}- l^2 V(\phi)=0
\ee
as a result the parameters of the theory are
\[
g= (\frac{2 n^2 l^2 e^{2l\phi}}{l^2\; V(\phi)-3 n^2 e^{2 l\phi}});\;\;\;
K=3\frac{n^2}{l^2}e^{2 l \phi}-V(\phi);
\]
\be
p=-2V(\phi)+\frac{3 n^2}{l^2}e^{2l\phi};\;\rho=\frac{3 n^2}{l^2}e^{2l\phi};
\ee
\[
w=1-\frac{2 l^2}{3 n^2}e^{-2l\phi}V(\phi).
\]
Now one can solve the set of equations (30), (31) and (32) to study the 
evolution of the phantom field, the scale factor and all the parameters of the 
theory under some suitable choice of the field potential $V(\phi)$, as 
before.\\

\noindent
{\underline{{Case 1}, Let, $V = V_{0}$ - a constant}}.\\

\noindent
Considering the potential to be a constant the action (1) can now be cast in 
the following form,
\be
S=\int d^4 x\sqrt{-g}[R+
\frac{n^2  l^2 e^{2l\phi}}{V_{0}\;l^2-3 n^2 e^{2 l\phi}}\phi,_{\mu}\phi^{,\mu}
-V_{0}],
\ee
Using equations (30) and (31), the field and the scale factor can now 
be evaluated as
\[
\phi=\frac{1}{l}\;log[\frac{n}{l}\sqrt{\frac{3}{V_{0}}}\;
tanh\{\sqrt{3 V_{0}}(t-t_{0})\}];
\]
\[
a= a_{0}\;cosh[\sqrt{3 V_{0}}(t-t_{0})]^{\frac{n^2}{V_{0}l^2}}.
\]
It is quite clear from the above expression that unless both the constants 
$n$ and $l$ are chosen negative, $\phi$ will not be well behaved. So we choose
$n=-m^2$ and $l=-\lambda^2$ and rewrite the above expressions as
\[
\phi=\frac{1}{\lambda^2}\;log[\frac{\lambda^2}{m^2}\sqrt{\frac{V_{0}}{3}}\;
coth\{\sqrt{3 V_{0}}(t-t_{0})\}];\]
\be
a= a_{0}\;cosh[\sqrt{3 V_{0}}(t-t_{0})]^{\frac{m^4}{V_{0}\lambda^4}}.
\ee
Rest of the parameters of the theory can as such be evaluated from equation 
(32) as,
\[
q= -[1+V_{0}coth^2\{\sqrt{3V_{0}}(t-t_{0})\}]+V_{0};\]
\[
g= \frac{6m^8 \lambda^4 tanh^2\{\sqrt{3V_{0}}(t-t_{0})\}}
{\lambda^8 V_{0}^2-9m^8 tanh^2\{\sqrt{3V_{0}}(t-t_{0})\}};\]
\be
K= \frac{9m^8}{\lambda^8 V_{0}}tanh^2\{\sqrt{3V_{0}}(t-t_{0})\}-V_{0};
\ee
\[
p=  \frac{9m^8}{\lambda^8 V_{0}}tanh^2\{\sqrt{3V_{0}}(t-t_{0})\}-2V_{0};
\]
\[
\rho= \frac{9m^8}{\lambda^8 V_{0}}tanh^2\{\sqrt{3V_{0}}(t-t_{0})\}; \]
\[
w= 1-\frac{2\lambda^8 V_{0}^2}{9m^8 tanh^2\{\sqrt{3V_{0}}(t-t_{0})\}}.
\]
Above set of solutions behave properly, provided, $V_{0}\lambda^4 > 3m^4$, as a 
result the field variable $\phi$ vanishes at the end, starting from an 
indefinite large value, while the scale factor grows exponentially. The 
deceleration parameter settles down to $-1$ starting from an indefinitely large 
value. The pressure starts from  $-2V_{0}$, gets reduced at the end while the 
energy-density starts from zero, increases but remains finite. The kinetic 
energy remains negative throughout the evolution, thus the dominant energy 
condition is violated all the time. The state parameter, starts from 
indefinitely large negative value and asymptotically comes close to, but less 
than $-1$, assuming, $3m^4$, though less than $V_{0}\lambda^4$, is pretty close 
to it. Note that, this is the only situation where $g$ could change sign 
provided $V_{0}\lambda^4 < 3m^4$, indicating a possible crossing of phantom 
divide line. However, this is not possible physically, since it makes the field 
negative. This solution is quite different from those obtained earlier with 
constant potential. Hence, the Hubble parameter plays crucial role in building 
solutions. \\  

\noindent
{\underline{{Case 2}, Let, $V = V_{0}e^{l\phi}$}}.\\

\noindent
Under the above choice of potential the action is expressed as
\be
S=\int d^4 x\sqrt{-g}[\frac{R}{2}+
\frac{n^2 l^2}{V_{0}l^2
e^{-l\phi}-3n^2}\phi,_{\mu}\phi^{,\mu}-V_{0}e^{l\phi}],
\ee
and equation (30) and (31) can be solved to obtain
\[
\phi=\frac{1}{l}log\left[\frac{V_{0}l^2}{3n^2+e^{-\frac{V_{0}l}{n}(t-t_{0})}}
\right];
\]
\be
a=a_{0}\sqrt[3][3n^2e^{\frac{V_{0}l}{n}(t-t_{0})}+1].
\ee
All other parameters of the theory (32) are
\[
q=-[1+\frac{e^{-\frac{V_{0}l}{n}(t-t_{0})}}{n^2}];
g=2n^2 l^2{e^{\frac{V_{0}l}{n}(t-t_{0})}};
\]
\be
p=-V_{0}^2 l^2 \left[\frac{3n^2+2
e^{-\frac{V_{0}l}{n}(t-t_{0})}}{(3n^2+e^{-\frac{V_{0}l}{n}(t-t_{0})})^2}\right];
\ee
\[
K=-\frac{V_{0}^2 l^2 
e^{-\frac{V_{0}l}{n}(t-t_{0})}}{(3n^2+e^{-\frac{V_{0}l}{n}(t-t_{0})})^2};
\rho= \frac{3V_{0}^2 l^2n^2}{(3n^2+e^{-\frac{V_{0}l}{n}(t-t_{0})})^2};
\]
\[
w=-[1+\frac{2}{3n^2}\;e^{-\frac{V_{0}l}{n}(t-t_{0})}].
\]
\begin{figure}
\includegraphics{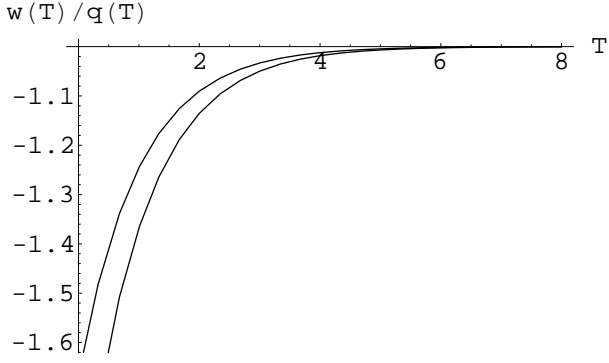}
\caption{The equation of state $w(T)$ (Upper curve) and the deceleration 
parameter $q(T)$ (lower curve), corresponding to solutions (38) are plotted 
against $T=\frac{V_{0}l}{n}(t-t_{0})$ for $n=1$. The graph shows how both 
touches the $-1$ line starting from below.
\label{figure2}}
\end{figure}

\begin{figure}
\includegraphics{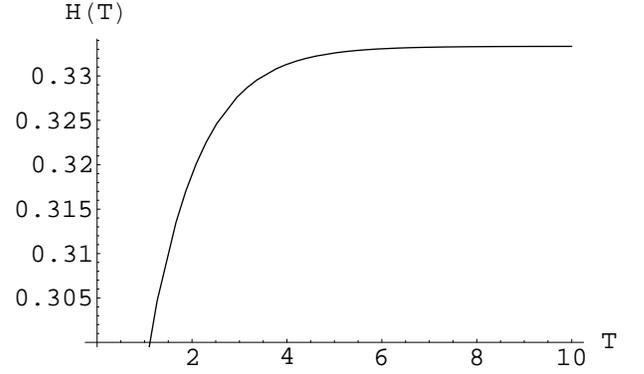}
\caption{The Hubble parameter $H(T)$, corresponding to solutions (38) has been 
plotted against same $T$ for $n=1, l=1,V_{0}=1$. It shows, by the time $w(T)$ 
and $q(T)$ touch $-1$ line, it also tends to a constant value starting from 
below with a positive constant value.
\label{figure3}}
\end{figure}
The above set of solutions (37) and (38), represents an ever accelerating 
universe which slows down to $q=-1$, while the increasing phantom field remains 
finite and well behaved at both ends, provided $V_{0}>\frac{3n^2+1}{l^2}$. The 
k-essence parameter increases indefinitely starting from a finite value, the 
pressure becomes less negative at the end, just as much positive as the energy 
density contributes, and as a result the kinetic energy vanishes, while the 
state parameter tends to $-1$, again indicating inflation at the end. Figure 2 
demonstrates how both the state and the deceleration parameters, touch the $-1$ 
line rather quickly, starting from large negative values. Figure 3 depicts the 
evolution of the Hubble parameter. The situation does not change if both $n$ 
and $l$ are considered negative, i.e., $g\dot\phi = -2s^2 e^{-l^2\phi}$ and 
$V=V_{0}e^{-m^2 \phi}$. However, $n$ can not be negative since it makes the 
field negative. Interestingly enough, if one considers, $l< 0$, particularly 
$l=-m^2$, the above set of solutions represent a collapsing Universe, whose 
scale factor reduces by a factor $(3n^2+1)^{\frac{1}{3}}$, while the field 
starting from a finite value provided, $V_{0}<\frac{3n^2+1}{l^2}$, grows 
indefinitely. Under such circmstances, the k-essence parameter, the energy 
density, the kinetic energy, the pressure all vanish at the end, while the 
deceleration parameter and the state parameter grow to indefinitely large 
negative value. So, it's a vacuum but different. Thus the positive and negative 
exonential potentials behave irreversibly.\\  
\section{Concluding remarks}
There exists a few exact solutions in the literature, corresponding to phantom 
fields, particularly in view of such generalized k-essence Lagrangian. Though 
most of the scientists agree that a single field Lagrangian is not viable to 
cross the phantom divide line \cite{pb}, yet there is an exception. It has been 
claimed  other way, by Andrianov et al \cite{aaa}. The cofusion can be removed 
only by presenting a class of exact solutions. Though we have presented a few, 
however it is possible to extract numerous solutions following the above 
method. The solutions with constant equation of state are found to represent 
all the phantom, quintessence and ordinary scalar field cosmological models, 
upon adjustment of a couple of parameters of the theory. Further, for 
$w\approx-1$, it is impossible to identify phantom with quintessence models. 
Crossing of the phantom divide line is indeed found to be impossible in such 
single field models, but all the solutions with dynamical state parameter are 
found to touch $-1$ line asymptotically and none of the solutions presented 
above suffer from instability or cosmological singularity like big rip. Thus we 
conclude in the spirit of \cite{av} and \cite{z} that phantom leads to 
inflation rather than big-rip.  
\begin{acknowledgments}
This work has been carried out during my visit at ICTP, Trieste, Italy. I would
like to thank ICTP for rendering all sorts of facilities and hospitality.
\end{acknowledgments}

\end{document}